# SMART CITIES: POTENTIALITIES AND CHALLENGES IN A CONTEXT OF SHARING ECONOMY

Ben Hur Monteiro Barizon, Federal University of Rio de Janeiro, Brazil, ben-hur.barizon@pped.ie.ufrj.br

Renata Lèbre La Rovere, Federal University of Rio de Janeiro, Brazil, renata@ie.ufrj.br

**Abstract:** The purpose of the present paper is to show how blockchain and IoT technologies can benefit smart city projects, which tend to spread in the context of the sharing economy. The article also aims to describe the challenges and potentialities of smart city projects. It was found that technology platforms can serve as a strategy to build the basis for product development (goods and services) and technology-based innovation.

## 1. INTRODUCTION

Smart cities are designed to create efficient and sustainable urban spaces that meet the demands of the citizens in an inclusive way using tools and technological management systems based on Information and Communication Technologies (ICT). The recent development of ICT, particularly blockchain systems, the Internet of things (IoT), and communication networks, can improve the quality of life, providing more efficient infrastructure and generating positive impact on other indicators related to the innovation processes described by ISO standard 37122.

The sharing economy is highly compatible with the smart city model. This is because the adoption of key ICT and of blockchain and IoT systems consolidates a new model of construction and renewal of urban spaces. Smart city applications are aimed at improving the quality of life, but the outcome of the implementation of this model also includes the generation of new businesses, which impacts the economy as well. (McKinsey & Company, 2019). Moreover, it should be noted that even though the sharing economy precedes the aforementioned technologies (La Rovere & Punzo, 2019), the advent of sharing platforms has accelerated the spread of this economic model.

The aim of the present article is to identify how smart city projects, which use blockchain technologies and IoT, can accelerate the development of a sharing economy in urban systems. Thus, we intend to contextualize and show how smart cities can restructure city life and improve urban centers with innovative goods services. To this end, section 2 describes the concept of sharing economy. In section 3, we will discuss how valuable blockchain and IoT technologies are for city projects. Section 4 presents the potentialities and challenges of smart city models in a context of social and economic transformations related to the sharing economy. In section 5, a conclusion and a synthesis of the results will be presented, as well as the implications for future studies.

## 2. SHARING ECONOMY

It is important to note that there is no consensus regarding the definition of the sharing economy. The difficulty to find a single definition for this phenomenon is to some extent due to its multiple interpretations and to the many concepts that are often associated with it, such as digital economy, gig economy, platform economy, peer economy, crowd economy, collaborative consumption, among many others. (Suciu et al., 2020)

La Rovere and Punzo (2019) defined the sharing economy as a phenomenon related to the growth of collaborative consumption. The main features of this type of consumption are: direct user-producer interaction; traceability, transparency and trust among agents; and customer participation. Tigre (2019) described the sharing economy as a recent phenomenon driven by the dissemination of online services and by widespread technology platforms. These innovative services seek to achieve hybrid synergy between the possession and the use of a product (goods or services). In other words, the sharing economy is a new business model aimed to meet the immediate and essential expectations of the user by using various resources that are idle or underused in a more dynamic way. Thus, the sharing economy aims to meet customers' needs while reducing new investments. This economic model is a good fit for a new type of consumer who values a more personalized experience focused on their identity and on their sense of community, without eliminating intermediaries from the value chain, but efficiently using new digital distribution models. The sharing economy is based on the development and application of ICT and on innovative services that can meet society's newest purchase and consumption habits. In that vein, smart city projects can accelerate the dissemination of innovative services related to the sharing economy.

In his studies on the importance of the factors of the sharing economy, Tigre (2019) states that the application of smart cities can include the technological platforms that are used in the sharing economy and that are the foundation of the development of products (goods and services). This new platform model comprises a set of standardized technologies and components, which indicates that sharing economy strategies may involve diverse ecosystems whose disruptive innovations promote new economic activities, which then can become new business models. These new business models may range from the selling of products to aggregate and more complex services focused on the customer experience, which, in turn, can promote disruptive models of customer service. These online platforms rely mainly on the technological capabilities made available through the Internet and seek to facilitate and optimize the resource allocation (connecting supply and demand) through ICT and data processing system infrastructures, creating new and innovative organizational models. Structures, assets and resources can be accessed directly on the aforementioned platforms or shared in a complementary and competitive way. The flexibility of the sharing economy represents an opportunity to achieve economies of scope, in which it is possible to adapt the supply to different activities, circumstances and to the needs of the demand.

A research by Davidson and Infranca (2016) identified that sharing has become an important feature of urban systems, since the society's needs and resources have increased. However, several implications are still not clear and transparent and, consequently, are not completely understood. In most large cities, sharing is a matter of necessity. Cars, bicycles, and even homes are shared, like numerous activities that are part of our daily lives. All of this was possible due to the extensive and intensive growth of communication technologies, positioning control technology (GPS), and to the widespread use of the Internet. Although new and innovative platforms are extensively studied by scholars from several research areas, there are still many fields to be explored in order to raise the understanding of how these platforms have changed the urban landscape, the social experience of living in cities, and even social interactions.

Curtis and Lehner (2019) showed that even without a precise definition, the sharing economy emphasizes the importance of ICT systems, which provides a means of communication and interconnection that can facilitate people's access to the products and services that drive this economy. The authors also mention how the sharing economy is compatible with smart cities' processes and models as the former increases efficiency, improves the quality of life and, above all, promotes effective cost reduction. On the other hand, as mentioned by Suciu et al. (2020), many authors have different opinions about the importance of the sharing economy when it comes to sustainability and viability. For instance, Botsman and Rogers (2011) praised the social impact of this economic model in urban spaces, since it promotes trust and social collaboration, whereas Belk (2010) points out that the use of idle capacity in the sharing economy can be good for the environment due to the reduction in energy consumption and also to the reduction of solid and liquid waste, actions that support sustainability within a community.

Other authors such as Frenken et al. (2015) and Kallis (2013) note that the sharing economy is a business model based on ICT that takes advantage of loopholes in regulatory norms, once it functions with unorthodox labor practices and it implements profitable or even offensive practices since, for the most part, workers are self-employed professionals who do not have any perks nor permanent contracts.

## 3. SMART CITIES, BLOCKCHAIN AND IoT

### 3.1. Blockchain

One of the main applications that can consolidate communication and exchange of information within the smart cities model using ICT, is available through the use of blockchain technology. This communication system, which is known worldwide by the virtual currency application base (cryptocurrencies), being bitcoin one of the best known examples, also represents a new communication system (transactions and proofs) in a safe way that can be implemented over the Internet and can even become a management model in which smart cities will be able to enjoy the synergy of this process in the form of an urban coordination service. The main objective is to use blockchain as a tool that allows the expansion of the reach and implementation of new technologies applied to smart cities.

Tigre (2019) describes blockchain as a chain of blocks of data or chained data. The author states that blockchain technology connects chunks of content in a process that resembles a fingerprint cross check. In other words, information is separated into different pieces, that is, the blocks. Each new block of data is stored, and it contains pieces of information from the previous block in the chain plus its own new content. Then, the new content is scanned and cross-checked. The process continues indefinitely as new data blocks generate the next "fingerprints". That is why the technology is called blockchain; these data blocks are somewhat chained to one another. The name refers quite literally to the architecture of the process. Blockchain technology aims to preserve trust in cryptographic transactions by avoiding the centralization of these data blocks. The technology also eliminates the disparities that users may come across in this type of transaction. Blockchains could be applied to various contexts and its benefits include the creation of smart contracts, and the promotion of self-pay and of more efficient decision-making in smart cities, so that the obstacles imposed by bureaucracy are removed and wealth generation becomes less complex for the society as a whole.

According to forecasts prepared by the American bank Merrill Lynch (2018), blockchain technology can assure that today's cities can be prepared to become, in the future, smart metropolises that will house a global population of 6.3 billion people by 2030. Additionally, it is worth mentioning that the blockchain technology generated global investments that amounted to nearly 1.3 trillion Euros in 2020.

To expand the scope of this technology in synergy with smart cities, a project called Blockchain4cities (BC4C) was created by the United Nations (UN) in 2016, which was initially a study group with representatives from different countries to control, integrate and coordinate various urban services more efficiently and transparently within urban management models. In this project a working group was created to identify examples of innovations in services within systems of digitization, urban management and sustainability, energy efficiency, mobility, citizen participation, waste treatment, among other activities. This study was attended by 26 specialists from several countries and four selected cities: Dubai (Middle East), Singapore (Asia), Chicago (America) and Gothenburg (Europe). The specialists were able to measure, among the main results of the research, the potential that blockchain technology has in of municipal governance, ensuring security, flexibility, in the safe transmission of information without intermediaries, under specific conditions applied to urban management problems and can serve as an example to be implemented on a larger scales to other countries around the world. The researchers within the Blockchain4Cities - BC4C (2016) project were able to measure some benefits and observe competitive advantages in the functioning of the cities that incorporated the blockchain system, which include: a) Connectivity and transparency => cities can generate interconnection with digital services (energy, mobility and security) using a real-time system that is open, single, transversal and accessible in the transmission of data to citizens; b) Direct Communication => cities can eliminate bureaucratic and time-consuming processes (notary offices and City Halls) allowing a direct communication without digital intermediaries between the public administration and citizens; c) Integrity of information => the use of technology allows encryption of information by sharing the data privately, securely and without risks of manipulation; d) Efficient management => the use of technology allows efficient control of resources in addition to ensuring citizen privacy in any public service.

Xie et al. (2019) is another study that highlights the potential that blockchain technology has of promoting the development and of improving services in smart cities. The study compared this technology with a database (shareable, immutable, decentralized and public). The authors then mention the main advantages of the blockchain system, which include: a) Decentralization => Blockchain systems normally work peer-to-peer without a centralized third party; b) Pseudonym: in the blockchain system, each node is linked to a pseudonymous public address, keeping real-world identities hidden. The inherent pseudonym is suitable for cases where the users' identities must be kept confidential; c) Transparency: Blockchain technology allows everyone to access all transaction records, which makes them transparent; d) Democracy: consensus algorithms are executed by all decentralized nodes to reach an agreement before a block is included in the blockchain. Thus, in the blockchain system, decisions are made by all nodes in a peer-to-peer way, which makes it more democratic; e) Security: In blockchain-based decentralized systems, it is rare to have a single point of failure. Thus, network security is improved; f) Immutability: In the blockchain system, all transactions are signed using digital signatures. In addition, data blocks are linked and protected through unilateral cryptographic hash functions. Any small modification generates a different hash and can be detected immediately, which makes the shared ledger immutable.

### 3.2. IoT (Internet of Things)

According to the description of the Internet Society (2015), the term Internet of things (IoT) is an emerging topic of technical, social and economic significance. It intends to cover, among other things, a universe of products (consumption, durable goods, cars and trucks, industrial and utility components, sensors and other everyday objects that will be connected to the Internet and have data analysis resources). IoT was designed to transform the way we work, live and play. The study estimated the impact of IoT on the Internet and on the economy, and the results are impressive. In some cases, the estimation projected up to 100 billion devices connected to the Internet, and a global economic impact of over $ 11 trillion by 2025. At the same time, as it happens with any projection, there are significant challenges that can constrain the unrealized potential of IoT. Some limitations include: cyber attacks to the devices connected to the Internet, surveillance issues, and the fear that the privacy of citizens and corporations may be at risk. Technically, it may be possible to restrict these effects, but there are still other political, legal, and personal challenges that will have to be addressed.

Regarding IoT, a study prepared by CISCO (2016) describes that the construction of an intelligent city is necessary, and the following aspects are indispensable in the process: a) Communication networks; (b) intelligent sensors; (c) mobile devices. All these components will have to be part of this great system that will be managed by a new technology called the Internet of things (IoT) that will be the backbone of the construction of smart cities, which will stimulate various initiatives that can improve and make the city more efficient and dynamic for all the people involved. Within this theme, this study shows that this model brings about guidelines in which new technologies called smart will have to deal with some of the problems crucial in urban planning, namely: a) Safety and Environment => it will be necessary to implement sensors that will monitor several different factors, from pollution control, water flow in the city to the verification and measurement of infrastructure conditions of bridges, streets, roads, signs, etc.; b) Buildings and Edifices => it will be necessary to plan and monitor the initial phases of new constructions and adaptation of the old ones considering the innovations that can have controlled energy efficiency (lighting , energy, gas emission, water flow, air conditioning, etc.); c) Urban Mobility => With the spiked urban population growth there will be a greater need that transportation is efficient. This quality control can be carried out by sensors that are monitored in real time, in order to have integrated solutions to handle heavy traffic and bottlenecks in urban systems (this includes the subway, bus, ferry, train, VLT and new technologies such as driverless cars and buses or people movers); d) Public Services => In an urban environment, we need to focus our energy on a synergy of options that can improve the way citizens live. Thus, a smart network project that manages the efficient use of energy and water is fundamental for the development and operation of the city.

### 3.3. Smart Cities

According to a report elaborated by the UN – World Urbanization Prospects (2014), by 2050 70% of the global population will be living in urban spaces, Therefore, it is necessary to build one or several models of smart cities that are inclusive and sustainable. The model of intelligent cities seeks to modify the dynamics and restructure the lives of the citizens, improving and promoting efficiency in the infrastructure of urban centers by supporting digital technologies. Blockchain technologies provide safer transactions and information and, in turn, by using IoT we will have more robust, comprehensive and resilient connection networks, facilitating the access to basic digital services that respond quickly and conveniently to the society's demands. With planning, structure, and strategic public policies, this model can be implemented effectively in any city that follows the required procedures and makes use of innovative and efficient technologies such as those used by blockchain and IoT. It is imperative to categorize the sectors involved in the application of smart cities, and the most promising areas to explore in the process. It is also important to explore the main applications of services focusing on the society's demands, and what lessons can be learned by using this model, comparing successful cases that have succeeded coupling public policies with technological and infrastructure input.

According to another study developed by the UN (United Nations) for their "2030 Agenda for Sustainable Development" initiative, the substantial population growth and the technological evolution have had exponential increase, especially during the 21st century. In this scenario, urban spaces and the society must adapt to globalization and to the migration phenomenon that occurs in many countries. Hence, it is indispensable that public and private managers have discussions on how to improve and adapt their urban and technological planning. In this context, smart cities are becoming more and more relevant.

A study conducted by the consulting firm McKinsey & Company (2019) showed that the intelligence of a city is not causally related to the installation of digital models and connections, basic infrastructure facilities, or to the strategic and digital operationalization of cities, but rather the correct use of all forms of technology as tools provide fast and accurate data and information analysis. According to the study, this is important so that public and private managers are able to make the best decisions to help the construction of processes that improve the quality of life of the population. The research presented the most important factors that can be listed as priorities to turn urban spaces into smart cities. Based on the results, it is possible to understand which indicators should be applied in the smart cities' model, as shown below::

**Figure 1 - Indicators of application of smart cities**

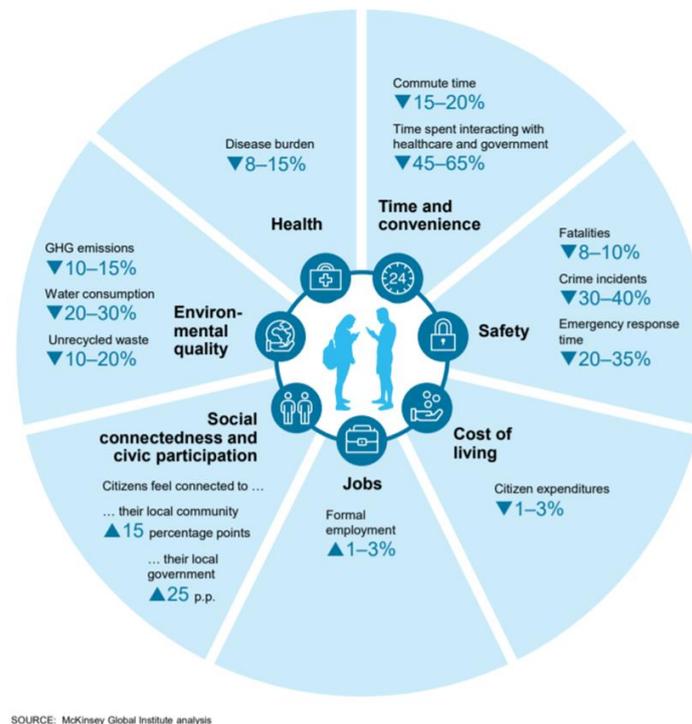

**Source: McKinsey Consulting**

The McKinsey & Company study (2019) found that there was an improvement in the quality of life, evidence shown in Figure 1, in which it is possible to see that the several models of implementation of ICT have the potential to improve key indicators by 10-30 percent.

## 4. POTENTIALITIES AND CHALLENGES OF SMART CITY MODELS

Suciu et al. (2020) show that a study developed by the European Commission (2020) defined smart cities as places where traditional networks and traditional services become more efficient through digital and telecommunications technologies to the benefit of residents, enterprises, and the society as a whole.

However, smart city projects face many challenges in their implementation. Xie et al. (2019), based on a study by Biswas and Muthukkumarasamy (2016), mapped out these challenges listing both technological and non-technological factors. Among the latter, it is possible to mention high financial investments and the availability of skilled labor. Technological factors include: a) Effective data treatment => the efficient collection and analysis of data, maintaining its reliability and integrity, promoting strategic planning that allows the provision of accessible public services and improves city management; b) Connection of devices in a decentralized way => in a decentralized system all nodes and connections can function in a flexible manner, adjusting each network to the dynamic scenario in which they are inserted, even with the increased number of electronic devices and sensors and complex applications, and with the increased volume of data; c) Participatory and transparent Public and Private Management => in this model government managers can disseminate information about city management (government affairs, environmental processes, decision-making) and companies (information about their clients) to meet the citizens' desire for participation, democracy and transparency; d) Efficient data sharing => efficient data sharing (IoT, organizational, personal) generates positive effects on the population, improving the supply of high-valued services, improving management, and promoting more accurate decision-making, because when the governments fail to share with organizations and individuals, it generates a lack of incentives and increases mistrust among agents.

A solution that could generate the necessary level of trust among agents, which is one of the pillars of the sharing economy, is the establishing of governance mechanisms from government institutions. According to Fernandez-Añes (2016), the aspects related to governance, sustainability and technological processes are considered by society to be the most important aspects in an intelligent city project. The establishment of transparent governance mechanisms are valuable to cities because they allow the provision of more efficient services that create change that favor the society, playing a fundamental role in improving the quality of life of the population.

To highlight the importance of governance services within smart city models as a public policy strategy, Mechant and Walravens (2018) show that these services have had a significant and substantial demand in the most recent years both in quantity of activities and in complexity. A study developed by the UN in 2014 showed an increase in demand for governance services by 46% within the countries of the European Union. Features such as ease of use, accessibility, trust value, and friendly services were pointed out, which is evidence of all the efforts governments have been making to provide services that meet the demands of their citizens.

Mechant and Walravens (2018) focused on the three main aspects of smart city projects: data, governance and participation. Concerning data, the authors stated that governments hold a wealth of information related to some aspects of a city and, consequently, of the citizens' lives. However, these data are not openly available, nor can they be easily interpreted, as they require mechanisms to process them. This lack of openness and transparency raised a movement that sought to have these data made available, structured, and machine-readable. The movement is now known as "open data", which gained significant strength in local and national governments. The second aspect is governance. The authors mentioned that new forms of governance and social innovation are related to the interactions between public managers and citizens, in which smart cities models are discussed and proposed. These interactions promote greater responsiveness of the public servants and more participative citizenship. This enhances the presence of governance mechanisms to provide synergies among governments, enterprises, Non-governmental Organizations (NGOs) and citizens. The third aspect is participation, in which the government explores new forms of collaboration and cooperation with the use of technology in a participatory design process (collection of principles and practices that include citizens as part of the design process of objects, services, spaces, and technologies), stimulating the creation of platforms that facilitate participation in lawmaking and policymaking.

A study by Suciu et al. (2020) pointed out the implications and strategies of the sharing economy in a smart city. In 2019, the authors had already identified that 54% of the world population lived in urban spaces. At the time, the degree of urbanization in Europe had already reached 74% of the total population of the continent. These projections corroborate the findings of a study conducted by the UN (2018) which estimated that over the course of 30 years the urban population worldwide would increase by 2.5 billion people, assuming geometric growth. In this context, it is worth mentioning that urban growth includes smaller cities and the so-called megacities. Hence, many are the challenges to be faced on all topics regarding the ISO standard 37120 (2014), including transportation, energy systems, housing, education, health, employability, among other types of infrastructure. In this regard, smart cities are designed to improve the quality of life of the population, turning cities into more inclusive and more sustainable urban spaces, stimulating the development of new business models that cross the borders between suppliers and consumers, aiming to reduce and restrict idle capacity and generating income for many new workers. Thus, the sharing economy model fits perfectly into this new global order.

Gori et al. (2015) describe that even with the current use of innovative platforms, the sharing economy has been growing for over 10 years. This has drawn the attention of organizations and institutions which have researched sharing economies in order to try to explain and also to evaluate their full potential. These studies show that smart cities and sharing economies have synergies as both promote innovative ideas that meet the needs of consumers to share resources, time, efficiency, skills and data first and foremost. Both smart cities and sharing economies rely on the support of ICT.

The scenario shown by Suciu et al. (2020) indicates that these new sharing technologies used in smart cities seek to change the way the citizens, the user or the customer, perceive access and consumption, depending on different types of resources. The authors also mention how they believe that new urban lifestyles are seen differently around the world. New purchase and consumption models transform the urban landscape, including the transportation, tourism, food, and retail industries that are creating new financial opportunities for the working class through the creation of new business models which represent a challenge for many traditional companies and institutions, forcing them to adapt in order to promote transparency to their consumers, thus paving the way and contributing to a more sustainable business culture.

## 5. FINAL REMARKS AND CONCLUSION

The present article showed that smart city projects can accelerate the trend of the current sharing economy, that is, the use of platforms for agents to interact freely and openly with transparency and, consequently, promoting higher levels of trust among agents.

It was seen in section 3 that new technologies such as blockchain and IoT are highly beneficial for smart city projects. However, as shown in section 4, the current context of the sharing economy requires data transparency and higher levels of trust among the agents who may not necessarily be present in these projects. Therefore, it is important to define governance of projects by establishing clear and open communication channels between the local government and the society through electronic government tools.

The use of blockchain technology and the IoT system in the context of the sharing economy will only be possible with the use of technology platforms that offer standardized preference services that optimize the existing resources in an innovative way by using ICTs. Moreover, these technologies stimulate the creation of more innovative and efficient organizational models, helping businesses adapt to provide new services that meet the society's demands and needs.

With strategic planning, investment in infrastructure, and public policies, smart city projects can be implemented effectively in any city that follows the required procedures. Therefore, it is important to categorize the sectors involved in the development of these cities to identify the most promising economic activities. Smart city projects can meet the society's wants and needs. In this sense, the guidelines for transparency of data are also important as they facilitate access to information on best practices and analysis of the possibilities of application of these practices to other cities.

Thus, it is important to note that smart city projects can only succeed if integration and commitment between public and private actors occur, allowing citizens to enjoy healthier and more sustainable lives. These projects can highly benefit from the blockchain technology and from interactions with the IoT system. Hence, this is an indication that investments made to favor the society can turn urban spaces into a universe in which everyone can share, take action, and contribute to promote sustainable development. In this context, knowledge generation will be indispensable for the growth and evolution of smart cities.

## REFERENCES AND CITATIONS

Agyeman, J., & Mclaren, D. (2014). "Smart Cities" Should Mean "Sharing Cities". Time Magazine. Retrieved January 20, 2021, from: < https://time.com/3446050/smart-cities-should-mean-sharing-cities/>.

Belk, R. (2010). Sharing. Journal of Consumer Culture, 36(5), 715-734. Retrieved January 20,2021, from: https://www.researchgate.net/publication/46553775_Sharing.

Biswas, K., & Muthukkumarasamy, V. (2016). Securing smart cities using blockchain technology. Proceedings IEEE HPCC/SmartCity/DSS. Retrieved January 20, 2021, from: < https://ieeexplore.ieee.org/document/7828539>.

Botsman, R., & Rogers, R. (2011). What's mine is yours: How collaborative consumption is changing the way we live. Collins.

Cisco. (2016). The Internet of Things. IBSG (Internet Business Solutions Group).

Curtis, S. K., & Lehner, M. (2019). Defining the sharing economy for sustainability. Sustainability, 11(3), 567. Retrieved Mach 22, 2021, from: <https://doi.org/10.3390/su11030567>.

Davidson, N. M., & Infranca, J. J. (2016). The Sharing Economy as an Urban Phenomenon. Yale Law & Policy Review, 34(2), 215-279. Retrieved March 22, 2021, from: https://ylpr.yale.edu/sharing-economy-urban-phenomenon.

European Commission. (2020). Smart Cities. Retrieved March 22, 2021, from: https://ec.europa.eu/info/eu-regional-           and-urban-development/topics/cities-and-urban-development/city-initiatives/smart- cities_en.

Evans, D. (2011). The Internet of Things. How the Next Evolution of the Internet Is Changing Everything. CISCO. IBSG (Internet Business Solutions Group).

Fernandez-Anez, V. (2016). Stakeholders approach to smart cities: A survey on smart city Definitions. Proceedings Smart-CT.

Frenken, K., Meelen, T., Arets, M., & Van de Glind, P. (2015). Smarter regulation for the sharing economy. The Guardian. Retrieved March 22, 2021, from: <http://www.theguardian.com/science/political-science/2015/may/20/smarter  regulation-for-the-sharing-economy>.

Getszko, D. (2020). Information and communication technologies in urban management: challenges for the mediation of smart cities. Information Center and Coordination of ponto BR.

Gori, P., Parcu, P. L., & Stasi, M. L. Smart Cities and Sharing Economy. European University Institute - Robert Schuman Centre for Advanced Studies. Retrieved March 21, 2021, from: http://cadmus.eui.eu/bitstream/handle/1814/38264/RSCAS_2015.

IBM. (2016). Device democracy: Saving the future of the Internet of Things. International Business Machines Corporation. Retrieved January 20, 2021, from: <http://www-935.ibm.com/services/us/gbs/thoughtleadership/internetofthings>.

Internet Society. (2015). The Internet of Things (IoT): An Overview. Internet Society. Retrieved March 20, 2021 from: https://www.internetsociety.org/resources/doc/2015/iot-overview.

ITU. (2020). United 4 Smart Sustainable Cities. ITU. Retrieved May 15, 2020, from: <https://www.itu.int/en/ITU-T/ssc/united/Pages/default.aspx>.

Kallis, G. (2013). AirBnb is a rental economy, not a sharing economy. The Press Project. Retrieved March 20, 2021, from: <http://www.thepressproject.net/article/68073/AirBnb-is-a-rental-economy-not-a- sharing-economy.

Lanza, J., Sanchez, L., Muñoz, L., Galache, J. A., Sotres, P., Santana, J. R., & Gutierrez, V.  et al. (2015). Large-scale mobile sensing enabled Internet-of-Things testbed for smart city services. International Journal of Distributed Sensor Network,11(8). Retrieved March 20, 2021, from: <https://doi.org/10.1155/2015/785061>.

La Rovere, R. L., & Punzo L. (2019). Inovações em turismo na Economia do Compartilhamento. In: P. Tigre, & A. Pinheiro. Inovação em Serviços na Economia do Compartilhamento. Saraiva.

Mckinsey & Company. (2018). Smart Cities: Digital Solutions for a more livable future. McKinsey Global Institute.

Mclaren, D., & Agyeman, J. (2015). Sharing Cities: A Case for Truly Smart and Sustainable Cities. MIT Press.

Mechant, P., & Walravens, N. (2018). E-Government and Smart Cities: Theoretical Reflections and Case Studies. Media and Communication, 6(4), 119–122.

Palmisano, S. J. (2008). A Smarter Planet: The Next Leadership Agenda. IBM, 6.

Suciu, M., Nasulea, D.F., & Nasulea, C. (2020). Smart City Innovation within the Sharing Economy: Urban Innovation and Collaborative Consumption. Proceedings of the 14th International Conference on Business Excellence, 14 (1), 1147-1157.

Sun, J., Yan, J., & Zhang, K. (2016). Blockchain-based Sharing services: What blockchain technology can contribute to smart cities. Spring Open, 26.

Tigre, P. B. (2019). Plataformas tecnológicas e a economia do compartilhamento. In: P. Tigre, A. Pinheiro. Inovação em Serviços na Economia do Compartilhamento. Saraiva.

Xie, J., Tang, H., Huang, T., Yu, F., Xie, R., Liu, J. & Liu, Y. (2019). Survey of Blockchain Technology Applied to Smart Cities: research issues and challenges. IEEE communications surveys & tutorials.